\documentclass[conference]{IEEEtran}
\IEEEoverridecommandlockouts
\usepackage{mypreamble} % NOTE: customize packages and newcommands
\usepackage{cite}
\usepackage{amsmath,amssymb,amsfonts}
\usepackage{mathtools}
\usepackage{algorithmic}
\usepackage{graphicx}
\usepackage{textcomp}
\usepackage{xcolor}
\usepackage{subfiles}
\def\BibTeX{{\rm B\kern-.05em{\sc i\kern-.025em b}\kern-.08em
    T\kern-.1667em\lower.7ex\hbox{E}\kern-.125emX}}

\begin{document}

\title{On The Multi-View Information Bottleneck Representation\\
}

\author{\IEEEauthorblockN{Teng-Hui Huang}
\IEEEauthorblockA{{Electrical and Computer Engineering} \\
{Purdue University}\\
West Lafayette, IN, USA \\
huan1456@purdue.edu}
\and
\IEEEauthorblockN{Aly El Gamal}
\IEEEauthorblockA{{Emerging Technologies Lab} \\
{InterDigital}\\
Los Altos, CA, USA \\
aly.elgamal@interdigital.com}
\and
\IEEEauthorblockN{Hesham El Gamal}
\IEEEauthorblockA{{Electrical and Information Engineering} \\
{The University of Sydney}\\
NSW, Australia \\
hesham.elgamal@sydney.edu.au}
%\and
%\and
%\and
%\and
}

\maketitle

\begin{abstract}
% a short summary of the recent interests in multiview data
In this work, we generalize the information bottleneck (IB) approach to the multi-view learning context. The exponentially growing complexity of the optimal representation motivates the development of two novel formulations with more favorable performance-complexity tradeoffs. The first approach is based on forming a stochastic consensus and is suited for scenarios with significant {\em representation overlap} between the different views. The second method, relying on incremental updates, is tailored for the other extreme scenario with minimal representation overlap. In both cases, we extend our earlier work on the alternating directional methods of multiplier (ADMM) solver and establish its convergence and scalability. Empirically, we find that the proposed methods outperform state-of-the-art approaches in multi-view classification problems under a broad range of modelling parameters.
\end{abstract}


\section{Introduction}
Recently, learning from multi-view data has drawn significant interest in the machine learning and data science community (e.g., \cite{sun2013survey,MvCsurvey2018,Federici2020Learning,wang2015deep,8471216}). In this context, a view of data is a description or observation about the source. For example, an object can be described in words or images. It is natural to expect learning from multi-view data to improve performance\cite{MvCG2019}. 

The challenges in multi-view learning are two-folded: First, as one can naively combine all views of observations to form one giant view which loses no information contained within them but would suffer from exponential growth of the dimensionality of the merged observation, we call this the \textit{performance-complexity trade-off}. Second, following this, if one instead opts to extract either view-shared or view-specific relevant features from each view, then heterogeneous forms of observations (e.g., audio and images) would make it difficult to learn low-complexity and meaningful representations with a unified method. In other words, the amount of \textit{representation overlap} across all view observations is important for efficient multi-view learning.

In addressing the challenges, several recent works have attempted to apply the IB \cite{tishby2000information} principle to multi-view learning for it matches the objective well, that is, trading-off relevance and complexity in extracting both view-shared and view-specific features \cite{Xu2014LMMVIB,4052777,DMIB2020,wang2019deep}. This generalization is known as the multi-view IB (MvIB) which is known to be a special case of the multi-terminal remote source coding problems with logarithmic loss. The logarithmic loss corresponds to soft reconstruction where the likelihood of all possible outcomes is received in contrast to a reconstructed symbol in conventional source coding problems~\cite{goldfeld2020information,8986754,8766879,estella2018distributed,zaidi2020information}.

In literature, the achievable region for the remote source coding problem is characterized in \cite{6651793}
 for discrete cases, and recently, jointly Gaussian cases as well \cite{8986754}. Along with the characterization, a variety of variational inference-based algorithms are proposed\cite{8766879,estella2018distributed}. This type of algorithms introduces extra variational variables to facilitate the optimization as by fixing one of the two sets of variables, the overall objective function is convex w.r.t. to the other set of variables. Then optimize in this alternating fashion, the convergence is assured.

Extending this line of research, our approach is rooted in a top-down information-theoretic formulation closely related to the optimal characterization of MvIB. Moreover, contrary to~\cite{wang2019deep} which relies on black-box deep neural networks, we propose two constructive information theoretic formulations with performance comparable to that of the optimal joint view approach. The two approaches are motivated by two extreme multi-view learning scenarios: The first is characterized by a significant representation overlap between the different views which favors our consensus-complement two stage formulation, whereas the second extreme scenario is characterized by a minimal representation overlap leading to our incremental update approach.

Different from existing variational inference-based algorithms that avoid dealing with the non-convexity of the overall objective function, in both of the proposed methods, we adopt the non-convex consensus ADMM as the main tool in deriving our solvers \cite{boyd2011distributed,attouch2009convergence,wang2019global}. These new solvers can, therefore, be viewed as generalizations of our earlier work on the single view ADMM algorithm~\cite{admmib2021}. More specifically, in the consensus-complement version, we separate the proposed Lagrangian into consensus and complement sub-objective functions and then proceed to solve the optimization problem in two steps. The new ADMM solver can hence efficiently form a consensus representation in large-scale multi-view learning problems with significantly lower dimensions compared to joint-view approaches. The same intuition is applied to the increment update approach as detailed in the sequel. %It is important to note that our two solvers are endowed with the ability to handle both random and deterministic representation mappings which distinguish them from existing greedy solvers. 
Finally, we prove the linear rate of convergence of our solvers under significantly milder constraints; as compared with earlier convergence results on this class of solvers~\cite{doi:10.1137/120878951,doi:10.1080/00207160.2016.1227432,NMTMA-14-438}. More specifically, we relax the need for a strongly convex sub-objective function and, moreover, establish the linear rate of convergence around a local optimal point in each case.   

\section{Multiview Information Bottleneck}\label{sec:mvib_formulation}
Given $V$ views with observations $\{X^{(i)}\}_{i=1}^V$ generated from a target variable $Y$, we aim to find a set of individual representations $\{Z\}$ that is most compressed w.r.t. the individual-view observations $X^{(i)},\forall i \in[V]$ but at the same time maximize their relevance toward the target $Y$ through $X^{(i)}$. 

Using a Lagrangian multiplier formulation, the problem can be casted as:
\begin{equation}\label{eq:lagrangian}
    \mathcal{L}_{\text{MvIB}}:= \gamma I(\{X\}_V;\{Z\})- I(Y;\{Z\}),
\end{equation}
where $\{X\}$ denotes the set of all $V$-views of observations and $\{Z\}$ is the set of representations to be designed. Note that if the observations are combined in this manner and treated as one view, the above reduces to the standard IB and can be solved with any existing single-view solver. However, combining all the observations in one giant view will result in an exponential increase in complexity (curse of dimensionality). 

A basic assumption in the multiview learning literature is the conditional independence \cite{10.1145/279943.279962,4052777}, where the observations of all views $\{X^{(i)}\}_{i=1}^V$ are independent given the target variable $Y$. That is, $p(\{x\}|y)=\prod_{i=1}^Vp(x^{(i)}|y)$. In the next two sections, we use this conditional independence assumption while constraining the set of allowable latent representations $\{Z\}$ to develop our two novel information-theoretic formulations of the Multi-view IB (MvIB) problem.

\subsection{Consensus-Complement Form}
Inspired by the co-training methods in multi-view research~\cite{10.1145/279943.279962}, we design the set of latent representations $\{Z\}$ to consist of a consensus representation $Z_c$ and view-specific complement components $Z_e^{(i)},\forall i\in[V]$. Then, by the chain rule of mutual information, the Lagrangian of \eqref{eq:lagrangian} becomes
\begin{multline}\label{eq:loss_cc_clean}
    \mathcal{L}_{\text{con}}=\gamma I(Z_c;\{X\})-I(Z_c;Y)\\
    +\sum_{i=1}^V\gamma I(Z_e^{(i)};\{X\}|Z_c,\{Z_e\}_{i-1})-I(Y;Z_e^{(i)}|Z_c,\{Z_e\}_{i-1}),
\end{multline}
where the sequence  $\{Z_e\}_{j}:=\{Z_e^{(1)},\cdots,Z_e^{(j)}\}$ is defined to represent the accumulated complement views.
To further simplify the above, we restrict the set of possible representations to satisfy the following conditions (similar to \cite{10.1145/279943.279962,wang2019deep}):
\begin{itemize}
    \item There always exist constants $\kappa_i,\forall i\in[V]$, independent of the observations $\{X\}$ such that $\kappa_iI(Z_c;X^{(i)})= I(Z_c;X^{(i)}|\{X\}_{i-1})$.
    \item $Y\rightarrow X^{(i)}\rightarrow Z_e^{(i)}\leftarrow Z_c$ forms a Markov chain. That is, $Z_c$ is side-information for $Z_e^{(i)}$.
    \item For each view $i\in[V]$, given the consensus $Z_c$, $\{Z_e\}$ are independent.
\end{itemize}

Under these constraints, \eqref{eq:loss_cc_clean} can be rewritten as the superposition of two parts, i.e., $\mathcal{L}:=\bar{\mathcal{L}}+\sum_{i=1}^V\mathcal{L}_e^{(i)}$, where the first component  $\bar{\mathcal{L}}$ is defined as the multi-view consensus IB Lagrangian:
\begin{equation}\label{eq:mv_lag_loss}
    \bar{\mathcal{L}}:=\sum_{i=1}^{V}\gamma_iI(Z_c;X^{(i)})-I(Z_c;Y),
\end{equation}
and the second consists of $V$ terms with each one corresponding to a complement sub-objective for each view:
\begin{equation}\label{eq:mv_lag_loss_cmpl}
    \mathcal{L}_e^{(i)}:=\gamma I(Z_e^{(i)};X^{(i)}|Z_c)-I(Z_e^{(i)};Y|Z_c),\forall i\in[V].
\end{equation}
We can now recast $\bar{\mathcal{L}}$ in \eqref{eq:mv_lag_loss} as:
\begin{equation}\label{eq:mv_lag_loss_vi}
    \begin{split}
        \bar{\mathcal{L}}:&=-\sum_{i=1}^V\gamma_iH(Z|X^{(i)})+\bigl(-1+\sum_{i=1}^V\gamma_i\bigr)H(Z)+H(Z|Y)\\
        &=\sum_{i=1}^VF_i(p_{z|x,i})+G(p_z,p_{z|y}).
    \end{split}
\end{equation}
Similarly, we rewrite \eqref{eq:mv_lag_loss_cmpl}, $\forall i\in[V]$, as:
\begin{multline}\label{eq:mv_alm_cmpl}
    \mathcal{L}_e^{(i)}=-\gamma H(Z_e^{(i)}|Z_c,X^{(i)})\\
    +(\gamma-1)H(Z_e^{(i)}|Z_c)
    +H(Z_e^{(i)}|Z_c,Y).
\end{multline}
By representing the discrete (conditional) probabilities as vectors/tensors, we can solve \eqref{eq:mv_lag_loss_vi} and \eqref{eq:mv_alm_cmpl} with augmented Lagrangian methods. We define the following vectors:
\begin{equation}\label{eq:def_pm_aug_vars}
\begin{split}
p_{z|x,i}&:=\begin{bmatrix}p(z_1|x_1^{(i)})\cdots  p(z_1|x^{(i)}_{N_i})\cdots p(z_L|x^{(i)}_{N_i})\end{bmatrix}^T,\\
p_z&:=\begin{bmatrix}p(z_1)&\cdots p(z_L)\end{bmatrix}^T,\\
p_{z|y}&:=\begin{bmatrix}p(z_1|y_1)\cdots p(z_1|y_T)\cdots p(z_L|y_K)\end{bmatrix}^T.
\end{split}
\end{equation}
where $N_i:=|\mathcal{X}^{(i)}|,\forall i\in[V],L:=|\mathcal{Z}|,K:=|\mathcal{Y}|$.
 For clarity, we rewrite the primal variables for each view as $p_{z|x,i}:=p_i$, and cascade the augmented variables which gives $q:=\begin{bmatrix}p_z^T&p_{z|y}^T\end{bmatrix}^T$. On the other hand, for the complement part, we define the following tensors:
\begin{equation}\label{def:tensor_mv_cmpl}
    \begin{split}
        \pi_{x,i}[m,n,r]&:=P(Z^{(i)}_e=z^{(i)}_{e,m}|Z_c=z_{c,n},X^{(i)}=x_r^{(i)}),\\
        \pi_{y,i}[m,n,r]&:=P(Z^{(i)}_e=z^{(i)}_{e,m}|Z_c=z_{c,n},Y=y_r),\\
        \pi_{z,i}[m,n]&:=P(Z^{(i)}_e=z^{(i)}_{e,m}|Z_c=z_{c,n}).
    \end{split}
\end{equation}

 Then we present the consensus-complement MvIB augmented Lagrangian as follows. For the consensus part:
\begin{multline}\label{eq:mv_aug_lag}
        \mathcal{L}_c(\{p_{i}\}_{i=1}^V,q,\{\nu_i\}_{i=1}^V)\\=
        \sum_{i=1}^V\left[F_i(p_i)+\langle \nu_i,A_ip_i-q\rangle+\frac{c}{2}\lVert A_ip_i-q\rVert^2\right]+G(q),
\end{multline}
where $\lVert\cdot\rVert$ is 2-norm. As for the complement part, let 
\begin{equation*}
\begin{split}
    F_{e,i}&=-\gamma H(Z_e^{(i)}|Z_c,X^{(i)}),\\ G_{e,i}&=(\gamma-1)H(Z_e^{(i)})+H(Z_e^{(i)}|Z_c,Y).
\end{split}    
\end{equation*}
As for the tensors used in the complement step, denote $\pi[t]$ the realization of the consensus representation $t\in\mathcal{Z}_c$, by Bayes' rule, we can recover the linear expression: $\pi_{y,i}[t]=\Lambda^{-1}_{z_c|y}[t]A_{x^{(i)}|y}^T\pi_{x,i}[t]$, where $\Lambda_{z_c|y}[t]$ is a diagonal matrix formed from the vector $p_{z_c=t|y},\forall y\in\mathcal{Y}$. To simplify notation, define the equivalent prior as a linear operator $\tilde{A}_{x^{(i)|y,t}}:=\Lambda^{-1}_{z_c|y}[t]A^T_{x^{(i)}|y}$, then we can express the augmented Lagrangian for the complement step as $\mathcal{L}^{(i)}_{e,c}=\sum_{t\in\mathcal{Z}_c}\mathcal{L}_{e,c}^{(i)}[t]$ and each term is defined as:
%FIXME:
%Define the cascaded tensor $\pi_{q,i}:=\begin{bmatrix}\pi_{z,i}^T&\pi_{y,i}^T\end{bmatrix}^T$, we get:
\begin{multline}\label{eq:mv_aug_lag_cmpl}
    \mathcal{L}_{e,c}^{(i)}[t]=F_{e,i}(\pi_{x,i}[t])+G_{e,i}(\pi_{q,i}[t])\\
    +\langle \mu_i,\tilde{A}_{e}\pi_{x,i}[t]-\pi_{q,i}[t]\rangle+\frac{c}{2}\lVert \tilde{A}_e\pi_{x,i}[t]-\pi_{q,i}[t]\rVert^2,
\end{multline}
where $c>0$ is the penalty coefficient and the linear penalty $A_ip_i-q$ for each view $i\in[V]$ encourages the variables $q$ and each $p_i$ to satisfy the marginal probability and Markov chain conditions. Specifically, let $\otimes$ denote the Kronecker product, $A_{z,i}:=I\otimes p_{x^{(i)}}^T,A_{z|y,i}:=I\otimes P_{x^{(i)}|y}^T$ where $P_{x^{(i)}|y}$ is the matrix form of the conditional distribution $p(x^{(i)}|y)$ with each entry $(m,n)$ equal to $p(x^{(i)}_m|y_n)$. 

%The tensors $A_e\pi_{x,i}-\pi_{z,i}$ can be vectorized to recover the linear form by considering a realization of the equivalent conditional variable:
% FIXME:
%\begin{equation*}
%    p(z_e^{(i)}|z_c,y)=\frac{\sum_{x^{(i)}}p(z_e^{(i)}|z_c,x^{(i)})p(z_c|x^{(i)})p(x^{(i)}|y)}{p(z_c|y_k)}.
%\end{equation*}
%In words, since $p(z_c|y)$ is given, for each $z_c\in\mathcal{Z}_c$ we can have the equivalent prior $p(x^{(i)}|y)p(z_c|x^{(i)})/p(z_c|y)$.

We propose a two-step algorithm to solve \eqref{eq:loss_cc_clean} described as follows. The first step is solving \eqref{eq:mv_aug_lag} through the following consensus ADMM algorithm. $\forall i\in[V]$:
\begin{subequations}\label{eq:main_alg_all}
\begin{align}
   p_{i}^{k+1} :=& \underset{p_{i}\in \Omega_{i}}{\arg\min}\,\mathcal{L}_c(\{p^{k+1}_{<i}\},p_{i},\{p^{k}_{>i}\} ,\{\nu_i^k\},q^k),\label{eq:main_alg_primal_i} \\
   \nu_i^{k+1}:=&\nu_i^k+c\left(A_ip_i^{k+1}-q^{k}\right),\label{eq:main_alg_dual}\\
   q^{k+1}:=&\underset{q\in\Omega_q}{\arg\min}\,\mathcal{L}_c(\{p^{k+1}_{i}\}_{i=1}^V,\{\nu_i^{k+1}\},q),\label{eq:main_alg_aug}
\end{align}
\end{subequations}
Then in the second step we solve  \eqref{eq:mv_aug_lag_cmpl} with two-block ADMM:
\begin{subequations}\label{eq:alg_cmpl_admm}
\begin{align}
        \pi^{k+1}_{e,i}&:=\underset{\pi_{x,i}\in\Pi^{(i)}_x}{\arg\min}\, \mathcal{L}_{e,c}(\pi_{x,i},\mu_i^k,\pi^k_{q,i}),\\
        \mu^{k+1}_{i}&:=\mu^{k}_i+c(\tilde{A}_{e,i}\pi^{k+1}_{x,i}-\pi^{k}_{q,i}),\\
        \pi^{k+1}_{y,i}&:=\underset{\pi_{y,i}\in\Pi^{(i)}_y}{\arg\min}\, \mathcal{L}_{e,c}(\pi^{k+1}_{x,i},\mu^{k+1}_i,\pi_{y,i}),
\end{align}
\end{subequations}

where in \eqref{eq:main_alg_all}, we use the short-hand notation  $\{p^{k+1}_{<i}\}:=\{p^{k+1}_{l}\}_{l=1}^{i-1}$ to denote the primal variables, up to $i-1$ views that are already updated to step $k+1$, and $\{p^k_{i<}\}:=\{p_{m}^k\}_{m=i+1}^{V}$ to denote the rest that are still at step $k$. We define $\{p^{k+1}_{<0}\}=\{\varnothing\}=\{p^k_{>V}\}$; in \eqref{eq:main_alg_all} and \eqref{eq:alg_cmpl_admm}, the superscript $k$ denotes the step index; each of $\Omega_i,\Omega_q,\Pi_x^{(i)},\Pi_y^{(i)}$ denotes a compound probability simplex. The algorithm starts with \eqref{eq:main_alg_primal_i}, updating each view in succession. Then the augmented variables are updated with \eqref{eq:main_alg_aug}. Finally, the difference between the primal and augmented variables are added to the dual variables \eqref{eq:main_alg_dual} to complete step $k$. After convergence of \eqref{eq:main_alg_all}, we run \eqref{eq:alg_cmpl_admm} in similar fashion for each view. And this completes the full algorithm.

\subsection{Incremental Update Form}
Intuitively, the consensus-complement form works well in the case where the common information in the observations $\{X\}$ across all views is abundant. However, if the views are almost distinct, where each view is a complement to the others, then the previous form will be inefficient in the sense that learning the common may have negligible benefit. To address this, we propose another formulation of the multi-view IB, by restricting the representation set to $\{Z^{(i)}\}_{i=1}^V$. The incremental update multi-view IB Lagrangian is therefore given by:
\begin{equation}\label{eq:mvib_inc}
    \mathcal{L}_{\text{inc}}:=\sum_{i=1}^V\gamma I(\{X\};Z^{(i)}|\{Z\}_{i-1})-I(Y;Z^{(i)}|\{Z\}_{i-1}).
\end{equation}
Again, to simplify the above, the incremental form will satisfy the following constraints:
\begin{itemize}
    \item For each view $i\in[V]$, the corresponding representation $Z^{(i)}$ only access $X^{(i)}$, so $Y\rightarrow X^{(i)}\rightarrow Z^{(i)}\leftarrow \{Z\}_{i-1}$ forms a Markov chain.
\end{itemize}
With the assumptions above, in each step, we can replace observations of all views $\{X\}$ with the view-specific observation $X^{(i)}$ and rewrite \eqref{eq:mvib_inc} as:
\begin{equation}\label{eq:inc_aug_lag}
    \mathcal{L}_{\text{inc}}:=\sum_{i=1}^V\gamma I(X^{(i)};Z^{(i)}|\{Z\}_{i-1})-I(Y;Z^{(i)}|\{Z\}_{i-1}).
\end{equation}
In solving \eqref{eq:inc_aug_lag}, we consider the following algorithm. At the~$i^{th}$ step, we have:
\begin{subequations}\label{eq:inc_alg}
\begin{align}
    &P^{(i)}_{z|x,z_{<i}}:=\underset{P\in\Omega^{(i)}}{\arg\min}\, \mathcal{L}_{\text{inc}}(P,\{P^{(j)}_{z|y,z_{<j}}\}_{j=1}^{i-1}),\label{eq:tensor_inc_def}\\
    &p(z^{(i)}|y,z_{<i})=\frac{\sum_{x_i}p(x^{(i)}|y)p(z^{(i)},z_{<i}|x^{(i)})}{p(z_{<i}|y)},
\end{align}
\end{subequations}
where $P^{(i)}_{z|x,z_{<i}}$ denotes the tensor form of a conditional probability $p(z^{(i)}|x^{(i)},z^{(i-1)},\cdots,z^{(1)}),\forall i\in[V]$. The tensor is the primal variable for step $i$ which belongs to a compound simplex $\Omega^{(i)}$. In the algorithm, for each step \eqref{eq:tensor_inc_def}, we solve it with \eqref{eq:main_alg_all} by setting $V=1$ and treating the estimators from the previous steps as priors. For example,  $p(z^{(2)},x^{(2)}|z^{(1)})=p(x^{(2)}|z^{(1)})p(z^{(2)}|z^{(1)},x^{(2)})$, and $p(x^{(2)}|z^{(1)})=\sum_yp(x^{(2)}|y)p(y|z^{(1)})$.
\section{Main Results}
We propose two new information-theoretic formulations of MvIB and develop optimal-bound achieving algorithms that are in parallel to existing solvers \cite{estella2018distributed,ugur2017general,8766879}; our main results are the convergence proofs for the proposed two algorithms. The convergence analysis goes beyond the MvIB and the recent non-convex multi-block ADMM convergence results as we further show that strong convexity on $\{F_i\}_{i=1}^V$ is not necessary for proving convergence \cite{NMTMA-14-438}. This new result connects our analysis to a more general class of functions that can be solved with multi-block non-convex consensus ADMM. For simplicity we denote the collective point at step $k$ as $w^k:=(\{p_i^k\},\{\nu^k_i\},q^k)$,  $\mathcal{L}_c^k:=\mathcal{L}_c(\{p_i^k\},\{\nu_i^k\},q^k)$ as the function value evaluated with~$w^k$ and $\mu_B,\lambda_B$ denote the smallest and largest singular value of a linear operator $B$ respectively,

%\begin{customthm}{1}\label{thm:informal_conv_mvc}
\begin{theorem}\label{thm:informal_conv_mvc}
Suppose $F_i(p_i)$ is $L_i$-smooth and $M_i$-Lipschitz continuous $\forall i\in[V]$ and $G(q)$ is $\sigma_G$-weakly convex. Further, let $\mathcal{L}_c$ be defined as in \eqref{eq:mv_aug_lag} and solved with the algorithm \eqref{eq:main_alg_all}. If the penalty coefficient satisfies $c>\max_{i\in[V]}\{(M_i\lambda_{A_i}L_i)/\mu_{A_iA_i^T},\sigma_G/V\}$, then the sequence $\{w^k\}_{k\in\mathbb{N}}$ is finite and bounded. Moreover, $\{w^k\}_{k\in\mathbb{N}}$ converges linearly to a stationary point $w^*$ around a neighborhood such that $\mathcal{L}_c^*<\mathcal{L}_c<\mathcal{L}_c^*+\delta$ for $\lVert w-w^*\rVert<\epsilon$ where $\delta,\epsilon>0$.
%\end{customthm}
\end{theorem}
\begin{IEEEproof}[proof sketch]
The details of the proof are deferred to Appendix \ref{appendix:conv_pf_all}. Here we explain the key ideas. 

Continuing on the proof sketch, the first step is to construct a sufficient decrease lemma (Lemma \ref{lemma:mvcib_suf_dec}) to assure that the function value $\mathcal{L}_c^k$ decreases from step $k$ to $k+1$ by an amount lower bounded by the positive squared norm $\lVert w^k-w^{k+1}\rVert^2$. We decompose $\mathcal{L}_c^k-\mathcal{L}_c^{k+1}$ according to each step of the algorithm \eqref{eq:main_alg_all}, $\forall i\in[V]$ as follows:
\begin{subequations}
\begin{align}
        {}&\mathcal{L}_c^k-\mathcal{L}_c^{k+1}\nonumber\\
        =&\begin{multlined}[t]
        \sum_{i=1}^V\left[\mathcal{L}_c(\{p^{k+1}_{<i}\},p^{k}_i,\{p^{k}_{> i}\},\{\nu^k\},q^k)\right.\\
        \left.-\mathcal{L}_c(\{p^{k+1}_{< i}\},p^{k+1}_i,\{p^k_{>i}\},\{\nu^k\},q^k)\vphantom{\mathcal{L}_c}\right]\end{multlined}\label{eq:subeq:fdiff_view}\\
        &+\begin{multlined}[t]
        \sum_{i=1}^V\left[\mathcal{L}_c(\{p^{k+1}\},\{\nu^{k+1}_{<i}\},\nu^k_{i},\{\nu^{k}_{> i}\},q^{k})\right.\\
        \left.-\mathcal{L}_c(\{p^{k+1}\},\{\nu^{k+1}_{< i}\},\nu^{k+1}_i,\{\nu^{k}_{>i}\},q^{k+1})\vphantom{\mathcal{L}_c}\right]
        \end{multlined}\label{eq:subeq:nudiff}\\
        &+\mathcal{L}_c(\{p^{k+1}\},\{\nu^k\},q^k)-\mathcal{L}_c(\{p^{k+1}\},\{\nu^k\},q^{k+1})\label{eq:subeq:qdiff}.
\end{align}
\end{subequations}
For each view, each difference in \eqref{eq:subeq:fdiff_view} can be lower bounded by using the convexity of $F_i$, and we get:
\begin{equation}\label{eq:sk_f_cvx_lb}
    \mathcal{L}_c(\{p_i^k\})-\mathcal{L}_c(\{p_i^{k+1}\})\geq \sum_{i=1}^V\frac{c}{2}\lVert A_ip_i^k-A_ip_i^{k+1}\rVert^2.
\end{equation}
On the other hand, in \eqref{eq:subeq:qdiff}, a similar lower bound for $G$ follows from its $\sigma_G$-weak convexity. This results in a negative squared norm $-\sigma_G/2\lVert q^k-q^{k+1}\rVert^2$. Nonetheless, by the first-order minimizer conditions \eqref{eq:cc_min_all} and the identity $2\langle u-v,w-u\rangle=|v-w|^2-|u-v|^2-|u-w|^2$, the negative term is balanced by the penalty coefficient $c$ as the corresponding lower bound is (with other variables fixed):
\begin{equation*}
    \mathcal{L}_c(q^k)-\mathcal{L}_c(q^{k+1})\geq \frac{cV-\sigma_G}{2}\lVert q^k-q^{k+1}\rVert^2.
\end{equation*}

As for the dual update,  \eqref{eq:subeq:nudiff} gives a combination of negative norms $-1/c\sum_{i=1}^V\lVert \nu_i^k-\nu_i^{k+1}\rVert^2$. It turns out that by the first-order minimizer condition of $F_i$ and its smoothness:
\begin{equation*}
    \nabla F_i(p_i^{k+1})=-A_i^T\nu_i^{k+1},
\end{equation*}
and that $A_i$ is full-row rank (holds for complement step):
\begin{equation}\label{eq:dual_grad_link}
    \lVert \nu_i^k-\nu_i^{k+1}\rVert^2
    \leq \frac{\lambda^2_{A_i}L_i^2}{\mu^2_{A_iA_i^T}}\lVert p_i^k-p_i^{k+1}\rVert^2,
\end{equation}
where $\mu_B,\lambda_B$ denote the smallest and largest singular value of a linear operator B.
 Then we need the following relation:
\begin{equation*}
    \lVert A_ip_i^k-A_ip_i^{k+1}\rVert\geq M \lVert p_i^k-p_i^{k+1}\rVert,\quad M>0,
\end{equation*}
which is non-trivial as $A_i$ is full-row rank. To address this, we adopt the sub-minimization path method in \cite{wang2019global}, which is applicable since $F_i$ is convex. Observe that \eqref{eq:main_alg_primal_i} is equivalent to a proximal operator:
\begin{equation*}
    \Psi_i(\eta):=\underset{\pi_i\in\Omega_i}{\arg\min}\,F(p_i)+\frac{c}{2}\lVert A_ip_i^k-\eta\rVert^2,
\end{equation*}
with $\eta:=A_ip_i^k$ at step $k$. Using this technique, we can have the desired result using the Lipschitz continuity of $F_i$:
\begin{multline*}
    \lVert\Psi_i(A_ip^k_i)-\Psi_i(A_ip_i^{k+1})\rVert=\lVert p_i^k-p_i^{k+1}\rVert\\
    \leq M_i\lVert A_ip_i^k-A_ip_i^{k+1}\rVert
\end{multline*}
prove the sufficient decrease lemma and hence the convergence (Appendix \ref{appendix:pf_main_thm}). 

We further prove that the rate of convergence is linear by explicitly showing the Kurdyka-{\L}ojasiewicz (K{\L} \cite{attouch2009convergence,li_pong_2018}) property is satisfied with an exponent $\theta=1/2$ (Appendix \ref{appendix:pf_expo_half}). This is based on the known result based on the K{\L} inequality, which characterizes the rate of convergence in to three regions in terms of $\theta$ \cite{attouch2009convergence} ($\theta=1/2$ corresponds to linear rate).
The proof for $\theta=1/2$ is again based on the convexity of $\{F_i\}$ and the weak convexity of $G$ and is referred to Appendix \ref{appendix:pf_expo_half}. We note that the linear rate holds around a neighborhood of a stationary point $w^*:=(\{p_i^*\},\{\nu_i^*\},q^*)$ which aligns with the results in \cite{li_pong_2018}.

%denote $c_{\min}$ the smallest $c$ that $\rho_q>0$.
%The key is to exploit lemma \ref{lemma:mvcib_suf_dec} and show that the function values $\bar{\mathcal{L}}_c(w^k)$ between consecutive steps are lower bounded $\bar{\mathcal{L}}_c(w^k)-\bar{\mathcal{L}}_c(w^*)\geq \rho^*\lVert w^k-w^*\rVert^2$. This exploits the fact that $G$ is $\sigma_G$-weakly convex, $L_q$-smooth and $F_i$ is convex. The convergence is assured by determining the conditions in terms of the penalty coefficient $c$ such that $\rho^*$ is always non-negative. 
%Lastly, the sufficient decrease lemma directly implies the rate of convergence as in the statement since:
%\begin{multline*}
%\lim_{K\rightarrow \infty}\bar{\mathcal{L}}_c^K-\bar{\mathcal{L}}_c^1\geq \rho^*\sum_{k=1}^K\lVert w^k-w^{k+1}\rVert^2\\
%\geq \rho^*K\min_{j\in[K]}\lVert w^j-w^{j+1}\rVert^2.
%\end{multline*}
%The l.h.s. is finite since mutual information are finite in discrete settings and by dividing the constants in the r.h.s. to the left, we complete the proof.
\end{IEEEproof}

As a remark, if the minimum element of a probability vector is bounded away from zero by a constant $\xi>0$, a commonly adopted smoothness condition in density and entropy estimation research \cite{6203416}, the sub-objectives $\{F_i\}_{i=1}^V$ and $G$ can be shown to be Lipschitz continuous and smooth. Furthermore, under smoothness conditions, $G$ is a weakly convex function w.r.t. $q$ (Lemma \ref{lemma:weak_cvx_g}). 
%Interestingly, the smoothness conditions are realized in the step-size selection process when implementing the proposed algorithms with gradient descent \cite{NocedalJorge2006No}, we leave the practical considerations for future exploration.

From Theorem \ref{thm:informal_conv_mvc}, the consensus-complement algorithm is convergent since the complement step is a special case of the algorithm \eqref{eq:main_alg_all} with $V=1$ while treating $p(z_c|x^{(i)})$ as an additional prior probability. The incremental algorithm is convergent following the same reason.

\begin{figure*}[t]
   \centerline{
        \subfloat[Classification Accuracy]{
           \includegraphics[width=2.9in]{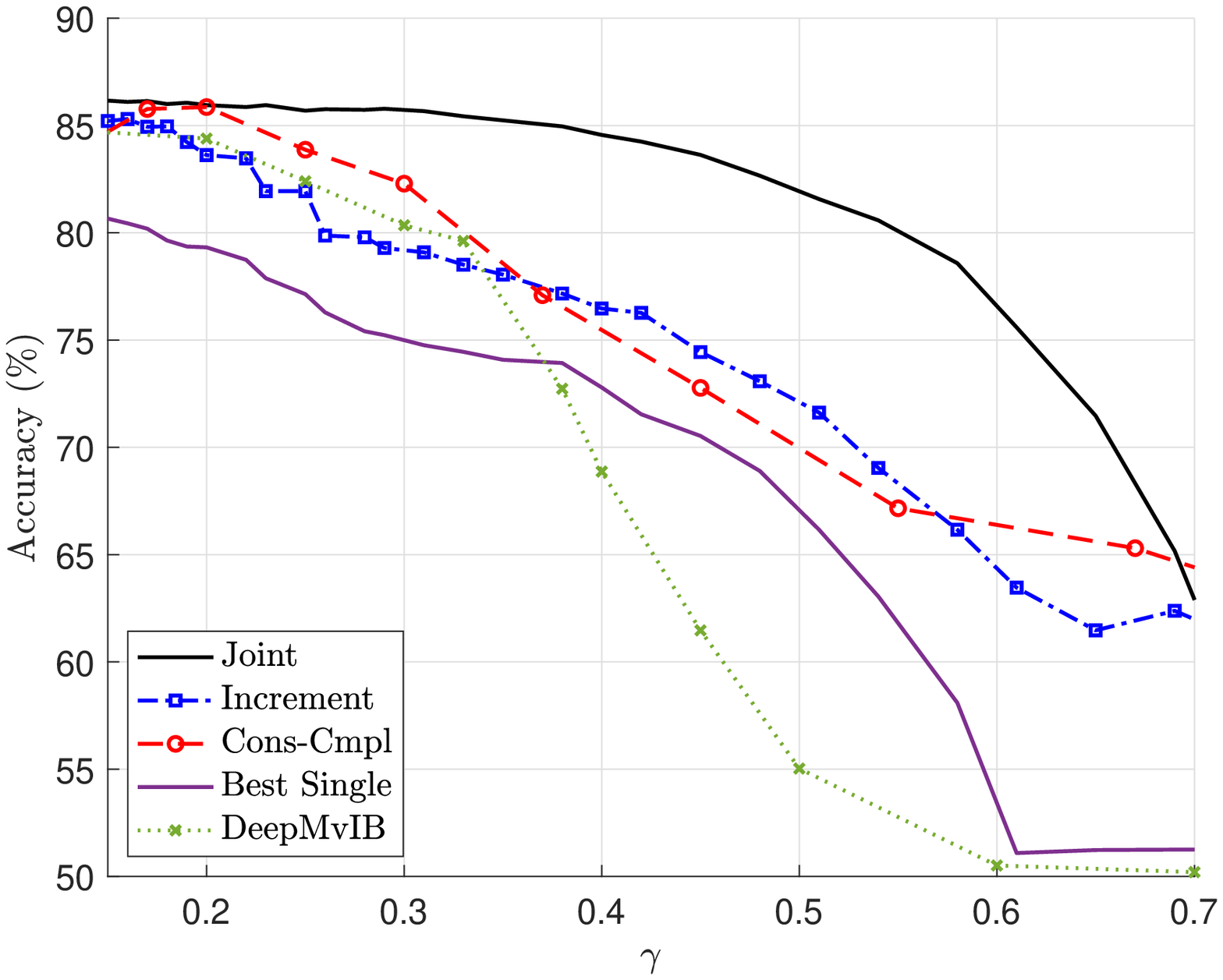}
           \label{subfig:bayes_acc}}
        \hfil
        \subfloat[Mutual information]{
          \includegraphics[width=2.9in]{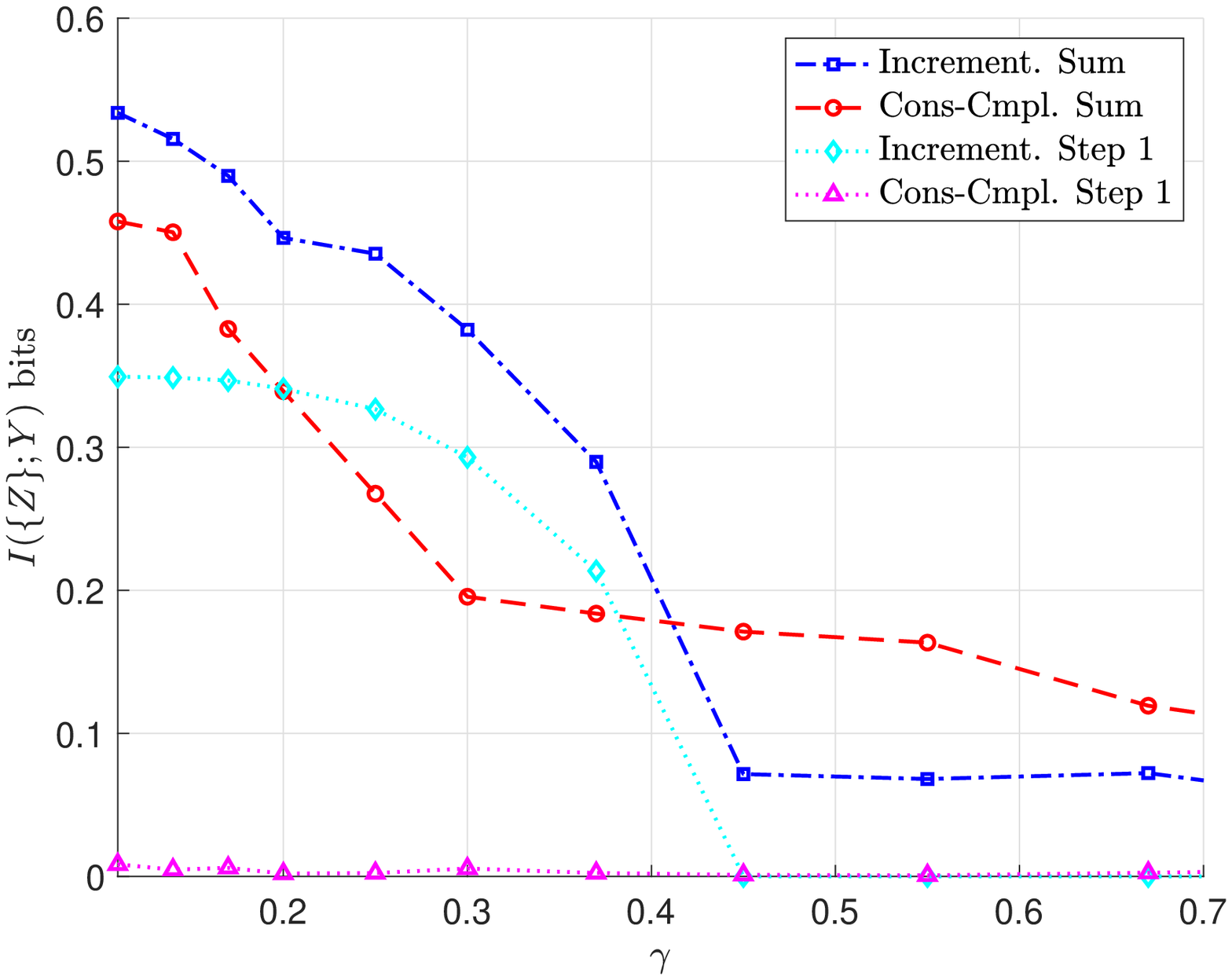}
           \label{subfig:mi_step1}
        }
   }
   \caption{Simulation results on synthetic datasets. We set $c=64,\min\{\varepsilon\}=10^{-11}$ and run the algorithms with random initialization. For simplicity, we let $\gamma_1=\gamma_2=\gamma$. The termination criterion for the proposed approaches is when the total variation (the linear constraints) between the primal and augmented variables $D_{TV}(A_ip_i||q)<10^{-6},\forall i\in [V]$ (convergent case), or the maximum number of iterations is reached (divergent case). Figure \ref{subfig:bayes_acc} follows the distribution given in \eqref{eq:eval_first_mat}  with \textit{Joint} denoting the joint-view IB; The distribution in Figure \ref{subfig:mi_step1} is given in \eqref{eq:eval_second_mat}. In both figures, the Bayes' decoder for \textit{Cons-Cmpl} is $
    p_{cc}(y|x^{(1)},x^{(2)})
    =\sum_{\{z_e^{(1)},z_e^{(2)},z_c\}}p(y|z_c,z_e^{(1)},z_e^{(2)})p(z_c|x^{(1)},x^{(2)})
    p(z_e^{(1)}|z_c,x^{(1)})p(z_e^{(2)}|z_c,x^{(2)})$ where we approximate $p(z_c=k|x^{(1)},x^{(2)})\approx p(z_c=k|x^{(1)})p(z_c=k|x^{(2)})/\sum_{t\in\mathcal{Z}_c}p(z_c=t|x^{(1)})p(z_c=t|x^{(2)})$;
On the other hand, for \textit{Increment}:
    $p_{inc}(y|x^{(1)},x^{(2)})=\sum_{\{z^{(1)},z^{(2)}\}}p(y|z^{(1)},z^{(2)})p(z^{(1)}|x^{(1)})p(z^{(2)}|z^{(1)},x^{(2)}).$}
   \label{fig:sim_all}
 \end{figure*}
\section{Numerical Results}
We evaluate the proposed two approaches for two-view, synthetic distributions.  For simplicity, we denote the consensus-complement approach as \textit{Cons-Cmpl} while the incremental update approach as \textit{Increment}. 

%We implement \eqref{eq:main_alg_all} with gradient descent similar to \cite{admmib2021} but in our case, a multi-block, consensus ADMM.

We simulate a classification task and compare the performance of the two proposed approaches to joint-view/single-view IB solvers \cite{1054855}, which are served as references for the best- and worst-case performance, and the state-of-the-art deep neural network-based method (DeepMvIB~\cite{8766879,wang2019deep}), with two layers of $4$-neuron, fully connected weights plus ReLU activation for each view. Given \eqref{eq:eval_first_mat}, we randomly sample $10000$ pairs of outcomes $(y,x^{(1)},x^{(2)})$ as testing data. Then we run the algorithms, sweeping through a range of $\gamma\in[0.1,0.7]$ and record the best accuracy 
from $50$ trials per $\gamma$. We use Bayes' decoder to predict the testing data, where we perform inverse transform sampling for the cumulative distribution of the decoders to obtain $\hat{y}$ for each pair of $(x^{(1)},x^{(2)})$. The data-generating distribution is:
\begin{equation}\label{eq:eval_first_mat}
    P(X^{(1)}|Y)=\begin{bmatrix}
        0.75 & 0.05\\
        0.20 & 0.20\\
        0.05 & 0.75
    \end{bmatrix},
    P(X^{(2)}|Y)=\begin{bmatrix}
        0.85 & 0.15\\
        0.15 & 0.85
    \end{bmatrix},
    % this is actually a counter example...
\end{equation}
with $P(Y)=\begin{bmatrix}0.5 & 0.5\end{bmatrix}^T$.
The result is shown in Figure \ref{subfig:bayes_acc}. The dimension of each of $Z_c,Z_e^{(2)},Z^{(2)}$ is $2$, and $3$ for each of $Z_e^{(1)},Z^{(1)}$. Clearly, the two proposed approaches can achieve comparable performance to that of the joint-view IB solver and outperform the deepMvIB over the range of $\gamma$ we simulated. Interestingly, \textit{Cons-Cmpl} outperforms \textit{Increment} in the best accuracy. This might be due to the abundance of representation overlap. To better investigate this observation, we further consider a different distribution with dimensions of all representations $|\mathcal{Z}_c|=|\mathcal{Z}_e^{(i)}|=|\mathcal{Z}^{(i)}|=3,\forall i\in\{1,2\}$:
\begin{gather}
    p(Y)=\begin{bmatrix}\frac{1}{3}&\frac{1}{3}&\frac{1}{3}\end{bmatrix}^T,P(X^{(1)}|Y):=\begin{bmatrix}
        0.90 & 0.20 & 0.20\\
        0.05 & 0.45 & 0.35\\
        0.05 & 0.35 & 0.45
    \end{bmatrix},\nonumber\\ P(X^{(2)}|Y):=\begin{bmatrix}
        0.25 & 0.10 & 0.55\\
        0.20 & 0.80 & 0.25\\
        0.55 & 0.10 & 0.20
    \end{bmatrix}.
    \label{eq:eval_second_mat}
\end{gather}
Observe that for each view in \eqref{eq:eval_second_mat}, there is one class ($y_1$ in view $1$ an $y_2$ in view $2$), that is easy to infer through $X^{(i)},i\in\{1,2\}$ while the remaining two are ambiguous. This results in low representation overlap and consensus is therefore difficult to form. In Figure \ref{subfig:mi_step1} we examine the components of the relevance rate $I(\{Z\};Y)$ where the \textit{Sum} is:
    $I(Z_c;Y)+\sum_{i=1}^2I(Z_e^{(i)};Y|Z_c))$
for \textit{Cons-Cmpl} and
    $I(Z^{(1)};Y)+I(Z^{(2)};Y|Z^{(1)})$
for \textit{Increment}. \textit{Step 1} indicates $I(Z_c;Y)$ in \textit{Cons-Cmpl}, and $I(Z^{(1)};Y)$ in \textit{Increment}. Observe that there is almost no increase in $I(Z_c;Y)$ over varying $\gamma$, and that \textit{Increment} has a greater relevance rate than \textit{Cons-Cmpl} when $\gamma<0.4$. Since it is known that the high relevance rate is related to high prediction accuracy \cite{shamir2010learning}, this example favors the \textit{Increment} approach as it is designed to increase the overall relevance rate view-by-view.

Lastly, we can compare the complexity of the two approaches in terms of the number of dimensions for the primal variables. For simplicity, let $|X|=|X^{(i)}|$,$|Z|=|Z_c|=|Z_e|=|Z^{(i)}|$. For \textit{Cons-Cmpl}, the number of dimensions for the variables scales as $\mathcal{O}(V|X||Z|^2)$ while for  \textit{Increment}, it grows as $\mathcal{O}(|X||Z|^V)$. The two methods both improve over the joint view as their complexity values scale as $\mathcal{O}(|X|^V|Z|)$ and $|Z|\ll|X|$ in general. Remarkably, the complexity for \textit{Cons-Cmpl} scales linearly in the number of views $V$ while we get an exponential growth with factor $|Z|$ for \textit{Increment}.
\section{Conclusion}
In this work, we propose two new information-theoretic formulations of MvIB and develop new optimal bound-achieving algorithms based on non-convex consensus ADMM, which are in parallel to existing solvers. We proposed two algorithms to solve the two forms respectively and prove their convergence and linear rates. Empirically, they achieve comparable performance to joint-view benchmarks and outperforms state-of-the-art deep neural networks-based approaches in some synthetic datasets. For future works, we plan to evaluate the two methods on available multiview datasets \cite{DBLP:journals/corr/abs-1902-03616,Cremers-Kolev-pami11} and generalize the proposed MvIB framework to continuous distributions \cite{franca2018admm}.

% FIXME: after finish revising, change this
\appendices
\section{Convergence Analysis}\label{appendix:conv_pf_all}
In this part, we prove the convergence of the consensus non-convex ADMM algorithm for the proposed two MvIB solvers, \eqref{eq:main_alg_all} and \eqref{eq:inc_alg}. Moreover, we demonstrate that the convergence rate is linear based on the recent non-convex ADMM convergence results through the K{\L} inequality. Specifically, we explicit show that the {\L}ojasiewicz exponents associate with the augmented Lagrangian for both forms is $1/2$ and therefore corresponds to linear rates. As mentioned in section \ref{sec:mvib_formulation}, the complement step and each level of the incremental update algorithms are special cases of the consensus step algorithm with a normalized linear operator for each realization of the conditioned representation and setting the numbers of view $V=1$, so it suffices to analyze the convergence and the associated rate of the non-convex consensus ADMM \eqref{eq:main_alg_all}. In solving \eqref{eq:main_alg_all}, we consider the following first-order optimization method and assume an exact solution exists and can be found at each step.

\subsection{Preliminaries}
% definition of Lojasweicz functions
% define K\L inequality
We first introduce the following definitions which allow us to study the properties of the sub-objective functions $F_i(p_{i}),G(q),\forall i\in[V]$.

We start with the elementary definitions of smoothness conditions for optimization.
\begin{definition}[Lipschitz continuity]
A real-value function $f:\mathbb{R}^n\mapsto\mathbb{R}$ is $M$-\textit{Lipschitz continuous} if $|f(x)-f(y)|\leq M|x-y|,M>0$ for $x,y\in\textit{\text{dom}}(f)$.
\end{definition}
A function is ``smooth'' if its gradient is Lipschitz continuous.
\begin{definition}[Smoothness]
A real-value function $f:\mathbb{R}^n\mapsto \mathbb{R},f\in\mathcal{C}^2$ is \textit{$L$-smooth} if $|\nabla f(x)-\nabla f(y)|\leq L|x-y|$, $L>0$ and $\forall x,y\in\textit{\text{dom}}(f)$.
\end{definition}
Note that if a \textit{$L$-smooth} function $f\in\mathcal{C}^2$, then the Lipschitz smoothness coefficient $L$ of $f$ satisfies $|\nabla^2f|\leq L$. In this work, the variables are cascade of (conditional) probability mass, and a common assumption in density/entropy estimation research is non-zero minimal measure \cite{6203416,han2020optimal}.
\begin{definition}[$\varepsilon$-infimal]\label{def:infimal}
A measure $f(x)$ is said to be \textit{~$\varepsilon$-infimal} if there exists $\varepsilon>0$ such that $\inf_{x\in \mathcal{X}}f(x)=\varepsilon$.
\end{definition}
Assuming \textit{$\varepsilon$-infimal} for a given set of primal variables $p_{i}$, we have the following results:
\begin{lemma}
Suppose $p_{i}$ in \eqref{eq:def_pm_aug_vars} is \textit{$\varepsilon_i$-infimal} $\forall i\in[V]$, $p_z$ is \textit{$\varepsilon_z$-infimal} and $p_{z|y}$ is \textit{$\varepsilon_{z|y}$-infimal}. Then $F_i$ is \textit{$\gamma_i/\varepsilon_i$-smooth} and $G$ is \textit{$1/\varepsilon_{q}$-smooth}. Where $\varepsilon_{q}:=\max\{|\sigma_z|/\varepsilon_{z},1/\varepsilon_{z|y}\}, \sigma_z:=1-\sum_{i=1}^V\gamma_i$.
\end{lemma}
\begin{IEEEproof}
For $F_i$, it suffices to consider a single view. Since $F\in\mathcal{C}^2$ and 
$\nabla F_i\leq\gamma_i\big[I\otimes\text{diag}(p_{x^{(i)}})\big]\big(\log{p_{z|x,i}}+\mathbf{1}\big)$, we have:
\[|\nabla^2 F_i|\leq\gamma_i\max_{x\in\mathcal{X}^{(i)},z\in\mathcal{Z}}\frac{p(x^{(i)})}{p(z|x^{(i)})}= \frac{\gamma_i}{\varepsilon_i}.\]
On the other hand, recall $q:=\begin{bmatrix}p_z &p_{z|y}\end{bmatrix}^T$, we can separate $G(q)$ into two parts, denotes $G(p_z),G(p_{z|y})$ respectively. For the first part $G(p_z)$:
\[
\nabla G(p_z)=\big(1-\sum_{i=1}^V\gamma_i\big)\big(\log{p_z}+\mathbf{1}\big)=\sigma_z\big(\log{p_z}+\mathbf{1}\big).
\]
On the other hand, for $p_{z|y}$:
\[
\nabla G(p_{z|y})=\big[I\otimes \text{diag}(p_y)\big]\big(\log{p_{z|y}}+\mathbf{1}\big).
\]
Since $G\in\mathcal{C}^2$, combine the two parts:
\begin{multline}
|\nabla^2G(q)|\\
\leq\max_{z\in\mathcal{Z},y\in\mathcal{Y}}\bigg\{\frac{|\sigma_z|}{p(z)},\frac{p(y)}{p(z|y)}\bigg\}\leq \max_{z\in\mathcal{Z},y\in\mathcal{Y}}\bigg\{ \frac{|\sigma_z|}{\varepsilon_z},\frac{1}{\varepsilon_{z|y}} \bigg\}.
\end{multline}
\end{IEEEproof}
Besides the smoothness conditions, given the joint probability $p(x,y)$, the (conditional) entropy functions are concave w.r.t. the associated probability mass \cite{Cover:2006:EIT:1146355}. The observation in \eqref{eq:mv_lag_loss} is that its non-convexity is due to a combination of differences of convex functions. By convexity, we refer to the following definition.
%If $L$-smooth function is also strongly convex, which is the case for $\varepsilon$-infimal negative entropy functions, then the following result applies.
\begin{definition}[Hypoconvexity]
A convex function $f:\mathbb{R}^n\mapsto\mathbb{R},f\in\mathcal{C}^2$ is \textit{$\sigma$-hypoconvex} if $\exists\sigma\in \mathbb{R},|\sigma|<+\infty$ such that $f(y)\geq f(x)+\langle\nabla f(x),y-x\rangle+\frac{\sigma}{2}|y-x|^2,\forall x,y\in \textit{\text{dom}}(f) $. In particular, if $\sigma>0$, $f$ is \textit{strongly convex}; if $\sigma<0$, $f$ is \textit{weakly convex}.
\end{definition}

In MvIB, given the $p_{x^{(i)}}$, it is easy to show that each $F_i,\forall i\in[V]$ is a convex function. On the other hand, for the function $G(q)$, if assume $\varepsilon_q$-infimal, we show in the following that it is weakly convex.
\begin{lemma}\label{lemma:weak_cvx_g}
Let $G(q)=\sigma_\gamma H(Z)+ H(Z|Y)$ and $q:=\begin{bmatrix}p_z&p_{z|y}\end{bmatrix}^T$. If $p_z,p_{z|y}$ is $\varepsilon_z,\varepsilon_{z|y}$-infimal measure respectively. Then the function $G$ is $\sigma_G$-weakly convex. Where $\sigma_\gamma:=1-\sum_{i=1}^V\gamma_i$ and $\sigma_G:=\max\{(2N_z|\sigma_\gamma|)/\varepsilon_z,(2N_zN_y)/\varepsilon_{z|y}\}$ with $N_z=|\mathcal{Z}|,N_y=|\mathcal{Y}|$.
\end{lemma}
\begin{IEEEproof}
see Appendix \ref{appendix:pf_weak_cvx_g}.
\end{IEEEproof}
From Lemma \ref{lemma:weak_cvx_g}, it turns out that the MvIB objective is a multi-block consists of $V$ convex $F_i$ in addition to a weakly convex $G$. This decomposition of the non-convexity of the overall objective enables us to generalize the recent strongly-weakly pair non-convex ADMM convergence results to consensus ADMM~\cite{guo_han_yuan_2017,jia_gao_cai_han_2021}. 

If a function satisfies the K{\L} properties, then its rate of convergence can be determined in terms of its {\L}ojasiewicz exponent.
\begin{definition}
A function $f(x):R^{|\mathcal{X}|}\mapsto R$ is said to satisfy the \textit{{\L}ojasweicz inequality} if there exists an exponent $\theta\in [0,1)$, $\delta>0$ and a critical point $x^*\in\Omega^*$ with a constant $C>0$, and a neighborhood $\lVert x-x^*\rVert\leq \varepsilon$ such that:
\[
\left|f(x)-f(x^*)\right|^{\theta}\leq C \text{dist}(0,\nabla f(x)).
\]
\end{definition}

In literature, there are a broad class of functions that are known to satisfy the K{\L} properties, in particular, the $o$-minimal structure (e.g., sub-analytic, semi-algebraic functions \cite{attouch2009convergence}).

% TODO:revise the definition 
% it is just a vague description right now.
\begin{definition}
A function $f(x):R^{|\mathcal{X}|}\mapsto R$ is said to satisfy the \textit{Kurdyka-{\L}ojasiewicz} inequality if 
there exists a neighborhood around $\bar{q}$ and a level set $Q:=\{q|q\in \Omega, f(q)<f(\bar{q})<f(q)+\eta\}$ with a margin $\eta>0$ and a continuous concave function $\varphi(s):[0,\eta)\rightarrow \mathbb{R}_+$, such that the following inequality holds:
\begin{equation}\label{eq:rate_kl_ineq}
    \varphi'(f(q)-f(\bar{q}))\textit{dist}(0,\partial f(q))\geq 1,
\end{equation}
where $\partial f$ denotes the sub-gradient of $f(\cdot)$ for non-smooth functions, and $\textit{dist}(y,A):=\inf_{x\in\mathcal{A}}\lVert x-y\rVert$ is defined as the distance of a set $A$ to a fixed point $y$ if exists. Note that if $\varphi(s):=\bar{c}s^{1-\theta}, \bar{c}>0$ then is recovers the definition of {\L}ojasiewicz inequality.
\end{definition}

The following elementary identity is useful in the convergence analysis, we list it for completeness.
\begin{equation}\label{eq:ident_three_pt}
    2\langle u-v,w-u\rangle=\lVert v-w\rVert^2-\lVert u-v\rVert^2-\lVert w-u\rVert^2.
\end{equation}
Lastly, by ``linear" rate of convergence, we refer to the definition in \cite{NocedalJorge2006No}:
\begin{definition}\label{def:rate_of_conv}
Let $\{w^k\}$ be a sequence in $\mathbb{R}^{n}$ that converges to a stationary point $w^*$ when $k>K_0\in\mathbb{N}$. If it converges \textit{$Q$-linearly}, then $\exists Q\in(0,1)$ such that
\[
\frac{\lVert w^{k+1}-w^*\rVert}{\lVert w^k-w^*\rVert}\leq Q,\quad \forall k>K_0.
\]
On the other hand, the convergence of the sequence is \textit{$R$-linear} if there is $Q$-linearly convergent sequence $\{\mu^k\},\forall k\in\mathbb{N}, \mu^k\geq 0$ such that:
\[
\lVert w^k-w^*\rVert\leq \mu^k, \quad \forall k\in\mathbb{N}.
\]
\end{definition}

%%%%%%%%%%
% Convergence Analysis
\subsection{Convergence and Rate Analysis}
As in the standard convex setting, to prove convergence of consensus ADMM, we simply need to establish a sufficient decrease lemma \cite{NesterovYurii2018Loco,boyd2011distributed}. However, since the MvIB problem is non-convex, the sub-objectives cannot be viewed as a monotone operator which leads to convergence naturally. Our key insight is that the non-convexity of the MvIB problem can be separated into a combination of a set of convex sub-objective $F_i$ and a single weakly-convex sub-objective $G$ which can be exploited to show convergence. This result requires certain smoothness conditions to be satisfied, which is a consequence when assuming $\varepsilon$-infimality (Definition \ref{def:infimal}) on the primal $p_i,\forall i\in[V]$ and the augmented variables $q$. With these smoothness conditions, it can be easily shown that $F_i$ is  $M_i$-Lipschitz continuous and $L_i$-smooth w.r.t. $p_i$ for some $M_i,L_i>0$. In addition to the properties of the sub-objective functions, it turns out that the structural advantages in consensus ADMM allows us to connect the dual update to the gradient of $F_i$, and we can therefore establish the desired results. Before presenting the results, we summarize the minimizer conditions as follows:
\begin{subequations}\label{eq:cc_min_all}
\begin{align}
  0&=\nabla F_i(p_i^{k+1})+A_i^T\left[\nu_i^k+ \left(A_ip_i^{k+1}-q^k\right)\right]\nonumber \\
  &=\nabla F_i(p_i^{k+1})+A_i^T\nu_i^{k+1},\label{eq:min_con_fi}\\
  \nu_i^{k+1}&=\nu_i^k+c\left(A_ip_i^{k+1}-q^k\right)\label{eq:min_con_dual},\\
  0&=\nabla G(q^{k+1})-\sum_{i=1}^V\nu_i^{k+1}-c\left(A_ip_i^{k+1}-q^{k+1}\right),\label{eq:min_con_g}%ERROR, not updated
\end{align}
\end{subequations}
where $i\in[V]$ denotes the view index. Following this, suppose there exists a stationary point $w^*:=(\{p_i^*\},\{\nu_i^*\},q^*)$ such that $\nabla \mathcal{L}_c(w^*)=0$, \eqref{eq:cc_min_all} reduces to:
\begin{equation}\label{eq:cc_min_stationary}
    A_ip_i^*=q^*,\quad \nabla F_i(p_i^*)=-A_i^Tp_i^*,\quad \nabla G(q^*)=\sum_{i=1}^V\nu_i^*.
\end{equation}
Furthermore, we impose the following set of assumptions to facilitate the convergence analysis:
\begin{assumption}\label{assump:conv}
\begin{itemize}
    \item There exists stationary points $w^*:=(\{p_{i}^*\},q^*,\{\nu_i^*\})$ that belong to a set $\Omega^*:=\{w|w\in\Omega,\nabla \mathcal{L}_c(w)=0\}$.
    \item $F_i(p_i),\forall i\in[V]$ is $L_i$-smooth, $M_i$-Lipschitz continuous and convex; $G(q)$ is $L_q$-smooth and $\sigma_G$-weakly convex.
    \item The penalty coefficient satisfies: \[c>\max_{i\in[V]}\{(M_i\lambda_{A_i}L_i)/\mu_{A_iA_i^T},\sigma_G/V\}.\]
\end{itemize}
\end{assumption}
With \eqref{eq:cc_min_all} and assumption \ref{assump:conv}, we can establish the sufficient decrease lemma, which is given below.
\begin{lemma} \label{lemma:mvcib_suf_dec}
Define $\mathcal{L}_c$ as in \eqref{eq:mv_aug_lag}. If Assumption \ref{assump:conv} is satisfied and the sequence $\{w^k\}_{k\in\mathbb{N}}$, with $w^k:=(\{p_i^k\},\{\nu_i^{l}\},q^k)$, is obtained from the algorithm \eqref{eq:main_alg_all}, then we have:
\begin{equation*}
    \mathcal{L}_c^k-\mathcal{L}_c^{k+1}\geq \sum_{i=1}^V\left[\frac{c}{2M_i^2}-\frac{\lambda^2_{A_i}L^2_i}{c\mu^2_{A_iA_i^T}}\right]\lVert p_i^{k+1}-p_i^k\rVert^2.
\end{equation*}
where $\mu_B,\lambda_B$ denote the smallest and largest singular value of a matrix $B$ respectively.
\end{lemma}
\begin{IEEEproof}
See Appendix \ref{appendix:pf_suf_dec_con}.
\end{IEEEproof}
As a remark, Lemma \ref{lemma:mvcib_suf_dec} implies the convergence of the non-convex ADMM-based algorithm depends on a sufficient large penalty coefficient $c$, and the minimum value that assures this, in turns, relies on the properties of the sub-objective functions $F_i,\forall i\in[V]$ and $G$. Note that both the Lipschitz continuity and the smoothness of $F_i$ are exploited to prove Lemma \ref{lemma:mvcib_suf_dec} which corresponds to the sub-minimization path method developed in \cite{wang2019global} as $F_i,\forall i\in[V]$ are convex, and the connection between dual update and the gradient of $F_i$ \cite{huang2022linearly,jia_gao_cai_han_2021}. 

In addition to convergence, it turns out that we can follow the recent results in optimization mathematics that adopt the K{\L} inequality to analyze the convergence and hence the rate of convergence for non-convex ADMM and prove that the proposed algorithms converge linearly. It is worth noting that the linear rate obtained through this framework is not uniform over the whole parameter space but converge linearly when the solution is located in the vicinity around local stationary points. In other words, the rate of convergence is locally linear. In the following we aim to use the K{\L} inequality to prove locally linear rate for the proposed two algorithms \eqref{eq:main_alg_all} and \eqref{eq:inc_alg}. As summarized in \cite{huang2022linearly}, three elements are needed to adopt the K{\L} inequality: 1) a sufficient decrease lemma. 2) Showing the {\L}ojasiewicz exponent $\theta$ of the objective function $\mathcal{L}$, solved with the algorithm to be analyzed is $1/2$ and 3) Contraction of the gradients $|\nabla \mathcal{L}^{k}|\leq C^*|w^k-w^{k-1}|,C^*>0$. Since we already have the first element, we can focus on the others. The desired result $\theta=1/2$ can be obtained through the following lemma.
\begin{lemma}\label{lemma:expo_half}
Let $\mathcal{L}_c$ be defined as in \eqref{eq:mv_aug_lag} and Assumption \ref{assump:conv} is satisfied. Let the sequence $\{w^k\}_{k\in\mathbb{N}}$ be obtained from the algorithm \ref{eq:main_alg_all} with $w^k:=(\{p^k_i\},\{\nu^k_i\},q^k)$ be the step $k$ collective point. Then the {\L}ojasiewicz exponent of $\mathcal{L}_c$ is $\theta=1/2$ around a neighborhood of a stationary point $w^*$ such that $|w-w^*|<\varepsilon$ and $\mathcal{L}_c(w^*)\leq\mathcal{L}_c(w)<\mathcal{L}_c(w^*)+\delta$ for $\varepsilon,\delta>0$.
\end{lemma}
\begin{IEEEproof}
See Appendix \ref{appendix:pf_expo_half}
\end{IEEEproof}
The element to adopt the K{\L} inequality is the contraction of the gradients of the augmented Lagrangian between consecutive updates.
\begin{lemma}\label{lemma:mvcc_grad_cont}
Let $\mathcal{L}_c$ be defined as in \eqref{eq:mv_aug_lag}. Suppose Assumption \ref{assump:conv} is satisfied and the sequence $\{w^k\}_{k\in\mathbb{N}}$ obtained through the algorithm \eqref{eq:main_alg_all} where $w^k:=(\{p_i^k\},\{\nu^k_i\},q)$ denotes the collective point at step $k$, then there exists a constant $C^*>0$ such that the following holds:
\begin{equation*}
    \lVert\nabla \mathcal{L}_c^k\rVert\leq C^*\lVert w^k-w^{k-1}\rVert.
\end{equation*}
\end{lemma}
\begin{IEEEproof}
See Appendix \ref{appendix:pf_grad_cont}.
\end{IEEEproof}
Then combining the lemmas, we can prove the locally linear rate of convergence of the non-convex consensus ADMM algorithm to form consensus MvIB representation. To be self-contained, the framework to adopt the K{\L} inequality is summarized in the following. Note that this result characterizes the rate of convergence into three regions in terms of the value of the {\L}ojasiewicz exponent and $\theta=1/2$ suffices in our case. For the complete characterization, we refer to \cite{attouch2009convergence,huang2022linearly} for details.
\begin{lemma}[Theorem 2 \cite{attouch2009convergence}]\label{lemma:kl_framework}
Assume that a function $\mathcal{L}_c(\{p_i\},\{\nu_i\},q)$ satisfies the K{\L} properties, define $w^k$ the collective point at step $k$, and let $\{w^k\}_{k\in\mathbb{N}}$ be a sequence generated by the algorithm \eqref{eq:main_alg_all}. Suppose $\{w^k\}_{k\in\mathbb{N}}$ is bounded and the following relation holds:
    \[\lVert\nabla \mathcal{L}_c^k\rVert\leq C^*\lVert w^k-w^{k-1}\rVert,\]
    where $\mathcal{L}_c^k:=\mathcal{L}_c(\{p_i^k\},\{\nu_i^k\},q)$ and $C^*>0$ is some constants. Denote the {\L}ojasiewicz exponent of $\mathcal{L}_c$ with $\{w^\infty\}$ as~$\theta$. Then the following holds:
    \begin{enumerate}[(i)]
        \item If $\theta=0$, the sequence $\{w^k\}_{k\in\mathbb{N}}$ converges in a finite number of steps,
        \item If $\theta\in(0,1/2]$ then there exist $\tau>0$ and $Q\in[0,1)$ such that
        \[|w^k-w^{\infty}|\leq \tau Q^k,\]
        \item If $\theta\in(1/2,1)$ then there exists $\tau>0$ such that
        \[|w^k-w^{\infty}|\leq \tau k^{-\frac{1-\theta}{2\theta-1}}.\]
    \end{enumerate}
\end{lemma}
Overall, the three elements for applying the K{\L} inequality are obtained from Lemma \ref{lemma:mvcib_suf_dec}, \ref{lemma:expo_half} and \ref{lemma:mvcc_grad_cont}. Then, by using Lemma \ref{lemma:kl_framework} we prove the linear rate of convergence.
\begin{customthm}{1}
Let $\mathcal{L}_c$ be defined as in \eqref{eq:main_alg_aug}, and Assumption \ref{assump:conv} is satisfied, then the sequence $\{w_k\}_{k\in\mathbb{N}}$ obtained through the algorithm \eqref{eq:main_alg_all} where $w_k:=(\{p^k_i\},\{\nu^k_i\},q)$ denote the collective point at step $k$ converges $Q$-linearly around a neighborhood of a stationary point $w^*$, satisfying $\lVert w-w^*\rVert<\varepsilon$ and $\mathcal{L}_c^*<\mathcal{L}_c<\mathcal{L}_c+\delta$ for some $\varepsilon,\delta>0$.
\end{customthm}
\begin{IEEEproof}
See Appendix \ref{appendix:pf_main_thm}.
\end{IEEEproof}

%%%%%%%%%%%%
% APPENDIX FOR PROOFS
%%%%%%%%%%%%%
\section{Proof of Lemma \ref{lemma:weak_cvx_g}}\label{appendix:pf_weak_cvx_g}
By definition, for two arbitrary $q^m,q^n\in\Omega_g$, $G(q^m)-G(q^n)$ consists of two parts. For the first part, we have:
\begin{equation*}
\begin{split}
    G(p_z^m)-G(p_z^n)&=(1-\sum_{i=1}^V\gamma_i)[H(Z^m)-H(Z^n)]\\
    &=\sigma_\gamma [H(Z^m)-H(Z^n)]
\end{split}
\end{equation*}
If $\sigma_\gamma <0$ then $G(p_z)$ is a scaled negative entropy function w.r.t. $p_z$ which is therefore $|\sigma_\gamma|$-strongly convex. In turns, the positive squared norm introduced by strong convexity is always greater than zero, which is $0$-weakly convex. On the other hand, if $\sigma_\gamma >0$, let $\sigma_\gamma=1$ without loss of generosity, we have:
\begin{equation}\label{eq:pf_lemma_weakcvx_pz}
\begin{split}
    {}&H(Z^m)-H(Z^n)\\
    =&\sum_{z}\langle p_z^m-p_z^n,-\log{p_z^n}\rangle-D_{KL}(p_z^m||p_z^n)\\
    \geq&\langle \nabla H(Z^n),p_z^m-p_z^n\rangle-\frac{1}{\varepsilon_z}\lVert p_z^m-p_z^n\rVert^2_1\\
    \geq&\langle \nabla H(Z^n),p_z^m-p_z^n\rangle-\frac{N_z}{\varepsilon_z}\lVert p_z^m-p_z^n\rVert^2_2
\end{split}
\end{equation}
where the first inequality is due to the reverse Pinsker's inequality \cite{DBLP:journals/corr/Sason15c} while the last inequality is due to norm bound: $\lVert x\rVert_1\leq \sqrt{N}\lVert x\rVert_2, x\in\mathbb{R}^n$. Then, for the second part, consider the following:
\begin{align*}
    {}&H(Z^m|Y)-H(Z^n|Y)\\
    &=\sum_yp(y)[\langle p^m_{z|Y}-p^n_{z|Y},-\log{p^m_{z|Y}}\rangle-D_{KL}(p^m_{z|Y}||p^n_{z|Y})]\\
    &\geq \langle \nabla H(Z^m|Y),p^m_{z|y}-p^n_{z|y}\rangle-E_y\bigl[\frac{1}{\epsilon_{z|y}}\lVert p^m_{z|Y}-p^n_{z|Y}\rVert_1^2\bigr]\\
    &\geq \langle \nabla H(Z^m|Y),p^m_{z|y}-p^n_{z|y}\rangle-\frac{N_zN_y}{\epsilon_{z|y}}\lVert p^m_{z|y}-p^n_{z|y}\rVert_2^2.
\end{align*}
where the first and second inequality follows the same reasons in \eqref{eq:pf_lemma_weakcvx_pz}. Combining the above discussions, we conclude that $G(q)$ is $\sigma_G$-weakly convex where:
\[\sigma_G:=\max\{\frac{2N_z|\sigma_\gamma|}{\varepsilon_z},\frac{N_zN_y}{\varepsilon_{z|y}}\}.\]

\section{Proof of Lemma \ref{lemma:mvcib_suf_dec} }\label{appendix:pf_suf_dec_con}
We divide the proof into three parts according to $p_i,\nu_i,q$ updates sequentially.

First, for each view $i\in[V]$, the $p_i$ updates from step $k$ to $k+1$ with $\{\nu_i^k\},q^k$ fixed, denote $\mathcal{L}_c(p_i^k)$:
\begin{equation}\label{eq:pf_suf_pi}
\begin{split}
    {}&\mathcal{L}_c(p_i^k)-\mathcal{L}_c(p_i^{k+1})\\
    =&\begin{multlined}[t]
    F_i(p_i^k)-F_i(p_i^{k+1})
    +\langle\nu_i^k,A_i(p_i^k-p_i^{k+1})\rangle\\
    +\frac{c}{2}\lVert A_ip_i^k-q^k\rVert^2
    -\frac{c}{2}\lVert A_ip_i^{k+1}-q^k\rVert^2
    \end{multlined}\\
    \geq& \begin{multlined}[t]
    \langle\nabla F_i(p_i^{k+1}),p_i^k-p_i^{k+1} \rangle +\langle\nu_i^k,A_i(p_i^k-p_i^{k+1})\rangle\\
    +\frac{c}{2}\lVert A_ip_i^k-q^k\rVert^2
    -\frac{c}{2}\lVert A_ip_i^{k+1}-q^k\rVert^2
    \end{multlined}\\
    =&\begin{multlined}[t]
    -c\langle A_ip_i^{k+1}-q^k,A_i(p_i^k-p_i^{k+1}) \rangle\\ +\frac{c}{2}\lVert A_ip_i^k-q^k\rVert^2
    -\frac{c}{2}\lVert A_ip_i^{k+1}-q^k\rVert^2
    \end{multlined}\\
    =&\frac{c}{2}\lVert A_ip_i^k-A_ip_i^{k+1}\rVert^2,
\end{split}
\end{equation}
where the inequality is due to the convexity of $F_i$ and the last two lines follow the minimizer condition \eqref{eq:min_con_fi} and the identity \eqref{eq:ident_three_pt} respectively.

Second, for the dual update of each view:
\begin{equation}\label{eq:pf_suf_nui}
\mathcal{L}_c(\nu_i^k)-\mathcal{L}_c(\nu_i^{k+1})=-\frac{1}{c}\lVert A_ip_i^{k+1}-q^k\rVert^2.
\end{equation}
Lastly, for the $q$ updates, from the $\sigma_G$-weak convexity of the sub-objective function $G$:
\begin{equation}\label{eq:pf_suf_q}
\begin{split}
    {}&\mathcal{L}_c(q^k)-\mathcal{L}_c(q^{k+1})\\
    =&\begin{multlined}[t]
    G(q^k)-G(q^{k+1})
    +\sum_{i=1}^V\left[\langle\nu_i^{k+1},q^{k+1}-q^k\rangle\right.\\
    \left.+\frac{c}{2}\lVert A_ip_i^{k+1}-q^k\rVert^2-\frac{c}{2}\lVert A_ip_i^{k+1}-q^{k+1}\rVert^2\vphantom{\langle\nu_i^{k+1}}\right]
    \end{multlined}\\
    \geq &\begin{multlined}[t]
    \langle\nabla G(q^{k+1}),q^k-q^{k+1}\rangle -\frac{\sigma_G}{2}\lVert q^k-q^{k+1}\rVert^2\\
    +\sum_{i=1}^V\left[\langle \nu_i^{k+1},q^{k+1}-q^k\rangle+\frac{c}{2}\lVert A_ip_i^{k+1}-q^k\rVert^2\right.\\
    \left.-\frac{c}{2}\lVert A_ip_i^{k+1}-q^{k+1}\rVert^2\vphantom{\langle\nu_i^{k+1}}\right]
    \end{multlined}\\
    =&\begin{multlined}[t]
    c\sum_{i=1}^V\left[\langle A_ip_i^{k+1}-q^{k+1},q^k-q^{k+1}\rangle \right.\\
    \left.+\frac{c}{2}\lVert A_ip_i^{k+1}-q^k\rVert^2-\frac{c}{2}\lVert A_ip_i^{k+1}-q^{k+1}\rVert^2\vphantom{\langle A_i}\right]\\
    \frac{-\sigma_G}{2}\lVert q^k-q^{k+1}\rVert^2.
    \end{multlined}\\
    &=\frac{cV-\sigma_G}{2}\lVert q^k-q^{k+1}\rVert^2. % ERROR, not updated
\end{split}
\end{equation}
Combining \eqref{eq:pf_suf_pi} \eqref{eq:pf_suf_nui} $\forall i\in[V]$ and \eqref{eq:pf_suf_q}, we have:
\begin{multline*}
    \mathcal{L}_c^k-\mathcal{L}_c^{k+1}\geq \frac{cV-\sigma_G}{2}\lVert q^k-q^{k+1}\rVert^2\\
    +\sum_{i=1}^V\left[\frac{c}{2}\lVert A_ip_i^{k}-A_ip_i^{k+1}\rVert^2-\frac{1}{c}\lVert \nu_i^k-\nu_i^{k+1}\rVert^2\vphantom{\frac{c}{2}}\right],% ERROR, not updated
\end{multline*}
where $\mathcal{L}_c^k$ denotes the augmented Lagrangian evaluated with step $k$ solution. The next step is to address the negative squared norm $\lVert \nu_i^k-\nu_i^{k+1}\rVert$. Since $A_i$ is full-row rank $\forall i\in[V]$, consider the following:
\begin{equation}\label{eq:pi_rowrank_property}
    \lVert (A_iA_i^T)^{-1}A_iA_i^T(\nu_i^k-\nu_i^{k+1})\rVert^2\leq \frac{\lambda_{A_i}^2}{\mu^2_{A_iA_i^T}}\lVert A_i^T(\nu_i^k-\nu_i^{k+1})\rVert^2,
\end{equation}
where $\mu_B,\lambda_B$ denotes the smallest and largest singular value of a linear operator $B$, we connect the gradient of $F_i$ to the dual update:
\begin{multline}\label{eq:fi_grad_dual_link}
    \lVert \nu_i^k-\nu_i^{k+1}\rVert^2\leq \frac{\lambda_{A_i}^2}{\mu^2_{A_iA_i^T}}\lVert \nabla F_i(p_i^k)-\nabla F_i(p_i^{k+1})\rVert^2\\
    \leq \frac{L_i^2\lambda^2_{A_i}}{\mu^2_{A_iA_i^T}}\lVert p_i^k-p_i^{k+1}\rVert^2,
\end{multline}
where the last inequality is due to the $L_i$-smoothness of $F_i$. After applying \eqref{eq:fi_grad_dual_link}, we need the relation:
\begin{equation*}
    \lVert A_ip_i^k-A_ip_i^{k+1}\rVert\geq M^* \lVert p_i^k-p_i^{k+1}\rVert,
\end{equation*}
% updated, not revised yet
for a constant $M^*>0$. Note that by definition, $A_i,\forall i\in[V]$ is full-row rank, hence the above relation is non-trivial. Fortunately, since each $F_i$ is convex w.r.t. $p_i$, which assures a unique minimizer, we can follow the sub-minimization technique recently developed in \cite{wang2019global} to establish the desired relation. By defining the following proximal operator:
\begin{equation*}
    \Psi_i(\eta):=\underset{p_i\in\Omega_i}{\arg\min}\, F_i(p_i)+\frac{c}{2}\lVert A_ip_i-\eta\rVert^2,
\end{equation*}
which coincides with the $p_i$ update, we have:
\begin{multline}\label{eq:fi_submin_path}
    \lVert\Psi_i(A_ip_i^k)-\Psi_i(A_ip_i^{k+1})\rVert=\lVert p_i^k-p_i^{k+1}\rVert\\
    \leq M_i\lVert A_ip_i^k-A_ip_i^{k+1}\rVert,%error, not updated
\end{multline}
where the last inequality is due to the Lipschitz continuity of $F_i$. Applying \eqref{eq:fi_submin_path} to \eqref{eq:pf_suf_pi} and we have:
\begin{multline*}
    \mathcal{L}_c^k-\mathcal{L}_c^{k+1}\geq \frac{cV-\sigma_G}{2}\lVert q^k-q^{k+1}\rVert^2\\
    +\sum_{i=1}^V\left[\frac{c}{2M_i^2}-\frac{\lambda_{A_i}^2L_i^2}{c\mu^2_{A_iA_i^T}}\right]\lVert p_i^k-p_i^{k+1}\rVert^2.
\end{multline*}
This completes the proof.
\section{Proof of Lemma \ref{lemma:expo_half}}\label{appendix:pf_expo_half}
The proof is divided into two parts. We first establish the following relation between a step $k+1$ solution and a stationary point $w^*:=(\{p_i^*\},\{\nu_i^*\},q^*)$:
\begin{equation}\label{eq:pf_ub_valdiff}
    \mathcal{L}_c^{k+1}-\mathcal{L}_c^*\leq \frac{c}{2}\sum_{i=1}^V\lVert A_ip_i^{k+1}-A_ip_i^*\rVert^2.
\end{equation}
This is accomplished by the following relations. Start from $\mathcal{L}_c^{k+1}-\mathcal{L}_c^*$, For $F_i,\forall i\in[V]$ differences, using the convexity and the minimizer condition \eqref{eq:min_con_fi}:
\begin{equation}\label{eq:pf_ub_pi}
    \begin{split}
        F_i(p_i^{k+1})-F_i(p_i^*)\leq -\langle \nu_i^{k+1},A_ip_i^{k+1}-A_ip^*_i\rangle,
    \end{split}
\end{equation}
where we use the reduction of the minimizer conditions at a stationary point \eqref{eq:cc_min_stationary} to have $A_ip_i^*=q^*,\forall i\in[V]$. As for the $G(q)$ difference:
\begin{multline}
    G(q^{k+1})-G(q^*)\leq \sum_{i=1}^V\langle \nu_i^{k+1},q^{k+1}-q^*\rangle\\ % ERROR not updated
    +c\sum_{i=1}^V\langle A_ip_i^{k+1}-q^{k+1},q^{k+1}-q^*\rangle+\frac{\sigma_G}{2}\lVert q^{k+1}-q^*\rVert^2.
\end{multline}
Note that by assumption $c>\sigma_G/V$. Therefore, by combining the two with the inner products where the dual variables $\nu_i^{k+1}$ associated with, we have the desired result \eqref{eq:pf_ub_valdiff}. The second part is to construct the following relation:
\begin{equation}\label{eq:pf_ub_grad_step}
    \lVert \nabla \mathcal{L}_c(w^{k+1})\rVert^2\geq \sum_{i=1}^V\left( c^2\mu_{A_iA_i^T}+1\right)\lVert A_ip_i^{k+1}-q^{k+1}\rVert^2,
\end{equation}
which is straightforward to show since:
\begin{equation}\label{eq:pf_ub_grad}
    \begin{split}
        {}&\nabla \mathcal{L}_c(w^{k+1})\\
        =&\begin{bmatrix}
        \nabla F_i(p_i^{k+1})+A_i^T\left[\nu_i^{k+1}+c\left(A_ip_i^{k+1}-q^{k+1}\right)\right]\\
        A_ip_i^{k+1}-q^{k+1}\\
        \nabla G(q^{k+1})-\sum_{i=1}^V\left[\nu_i^{k+1}+c\left(A_ip_i^{k+1}-q^{k+1}\right)\right]
        \end{bmatrix}\\
        =&\begin{bmatrix}
        cA_i^T\left(A_ip_i^{k+1}-q^{k+1}\right)\\
        A_ip_i^{k+1}-q^{k+1}\\
        0
        \end{bmatrix},
    \end{split}
\end{equation}
where the last equality is due to the minimizer conditions \eqref{eq:cc_min_all}. Note that in \eqref{eq:pf_ub_grad}, for simplicity, we derive the relation considering a single $i\in[V]$ for $p_i$ and $\nu_i$.
With \eqref{eq:pf_ub_valdiff} and \eqref{eq:pf_ub_grad_step}, consider a neighborhood around a stationary point $w^*$ such that $|\bar{w}-w^*|<\varepsilon$ and $\mathcal{L}^*_c<\bar{\mathcal{L}}_c<\mathcal{L}^*_c+\delta$ for some $\varepsilon,\delta>0$, then we have:
\begin{equation}\label{eq:pf_expo_half_all}
    \begin{split}
        {}&\mathcal{L}^{k+1}_c-\mathcal{L}_c^*\\
        \leq&\begin{multlined}[t] \sum_{i=1}^V\left[\frac{c}{2}\lVert A_ip_i^{k+1}-A_ip_i^{*}\rVert^2\right.\\
        \left.+(c^2\mu_{A_iA_i^T}+1)\lVert A_ip_i^{k+1}-q^{k+1}\rVert^2\vphantom{\frac{c}{2}}\right]
        \end{multlined}\\
        \leq&\sum_{i=1}^V\frac{c}{2}\lVert A_ip_i^{k+1}-A_ip_i^*\rVert^2+\lVert \nabla \mathcal{L}_c^{k+1}\rVert^2\\
        \leq&\lVert\nabla\mathcal{L}_c^{k+1}\rVert^2\left(1+\frac{c\varepsilon^2}{2\lVert\nabla \mathcal{L}_c^{k+1}\rVert^2}\right)\\
        \leq &\lVert\nabla \mathcal{L}_c^{k+1}\rVert^2\left(1+\frac{c\varepsilon^2}{2\eta^2}\right),
    \end{split}
\end{equation}
where in the last equality, by construction, because $\mathcal{L}_c^{k+1}$ is not a stationary point, there exists a $\eta>0$ such that $|\nabla \mathcal{L}_c^{k+1}|>\eta$ \cite[Lemma 2.1]{li_pong_2018}.  Then we complete the proof as \eqref{eq:pf_expo_half_all} implies that the {\L}ojasiewicz exponent $\theta=1/2$.

\section{Proof of Lemma \ref{lemma:mvcc_grad_cont}}\label{appendix:pf_grad_cont}
From \eqref{eq:pf_ub_grad}, we have:
\begin{equation}
\begin{split}
    {}&\lVert\nabla \mathcal{L}_c^k\rVert^2\\
    \leq& \sum_{i=1}^V\left(c^2\lambda_{A_iA_i^T}+1\right)\lVert A_ip_i^k-q^k\rVert^2\\
    \leq& 2\sum_{i=1}^V\left(c^2\lambda_{A_iA_i^T}+1\right)\left[\lVert A_ip_i^k-q^{k-1}\rVert^2+\lVert q^k-q^{k-1}\rVert^2\right]\\
    =&\begin{multlined}[t]
    2\sum_{i=1}^V\left[ \left(\lambda_{A_iA_i^T}+\frac{1}{c^2}\right)\lVert \nu_i^k-\nu_i^{k-1}\rVert^2\right.\\
    \left.+\left(c^2\lambda_{A_iA_i^T}+1\right)\lVert q^k-q^{k-1}\rVert^2\vphantom{\lambda_{A_iA_i^T}}\right]
    \end{multlined}\\
    \leq&\begin{multlined}[t]
    2\sum_{i=1}^V\left(\lambda_{A_iA_i^T}+\frac{1}{c^2}\right)\left[\lVert\nu_i^k-\nu_i^{k-1}\rVert^2+\lVert p_i^k-p_i^{k-1}\rVert^2\right]\\
    +2\sum_{i=1}^V\left(c^2\lambda_{A_iA_i^T}+1\right)\lVert q^k-q^{k-1}\rVert^2,
    \end{multlined}
\end{split}
\end{equation}
Then there exists a positive constant $C^*>0$ such that:
\begin{equation*}
    \lVert\nabla\mathcal{L}_c^k\rVert\leq C^*\lVert w^k-w^{k-1}\rVert,
\end{equation*}
where $w^k:=(\{p_i^k\},\{\nu_i^k\},q^k)$ denotes the step $k$ collective point.
\section{Proof of Theorem \ref{thm:informal_conv_mvc}}\label{appendix:pf_main_thm}
We first show the convergence. By Assumption \ref{assump:conv}, the sufficient decrease lemma (Lemma \ref{lemma:mvcib_suf_dec}) holds. Consider a sequence $\{w_k\}_{k\in\mathbb{N}}$ obtained through the algorithm \eqref{eq:main_alg_all}, by sufficient decrease lemma, there exists some constants $\rho_{p,i},\rho_q>0,\forall i\in[V]$ such that:
\begin{equation}\label{eq:pf_thm_conv}
    \mathcal{L}_c^0-\mathcal{L}_c^\infty\geq  \sum_{k=0}^\infty\left[\sum_{i=1}^V\rho_{p,i}\lVert p_i^k-p_i^{k+1}\rVert^2+\rho_q\lVert q^k-q^{k+1}\rVert^2\right].
\end{equation}
In discrete settings, the $\mathcal{L}_c$ is lower semi-continuous and therefore the l.h.s. of \eqref{eq:pf_thm_conv} is bounded. Following this, observe that the r.h.s. of \eqref{eq:pf_thm_conv} is a Cauchy sequence, and hence $\forall i\in[V], p_i^k\rightarrow p_i^*,q^k\rightarrow q^*$ as $k\rightarrow \infty$. Given these, by \eqref{eq:fi_grad_dual_link}, this result implies $\nu_i^k\rightarrow \nu_i^*$ as $k\rightarrow \infty$. And we prove the convergence.

Given convergence, along with Lemma \ref{lemma:expo_half} and Lemma \ref{lemma:mvcc_grad_cont}, that is, the K{\L} property is satisfied with a {\L}ojasiewicz exponent is $\theta=1/2$, and the contraction of the gradient of $\mathcal{L}_c$ is established, by Lemma \ref{lemma:kl_framework}, we prove the rate of convergence of the sequence $\{w^k\}_{k\in\mathbb{N}}$ obtained through the algorithm \eqref{eq:main_alg_all} is $Q$-linear around a neighborhood of a stationary point $w^*$.
\bibliographystyle{IEEEtran}
\bibliography{references}

\end{document}